\documentclass[10pt,journal,compsoc, cspaper, twoside]{IEEEtran}

\IEEEoverridecommandlockouts

\usepackage[utf8]{inputenc}
\usepackage{wrapfig}

\usepackage{graphicx}

\usepackage{tcolorbox,fancyvrb,xcolor,tikz}

\usepackage{cite}
\usepackage{amsmath,amssymb,amsfonts}
\usepackage{algorithmic}
\usepackage{textcomp}
\usepackage{xcolor}
\usepackage{booktabs}
\usepackage{multirow}
\usepackage{subcaption}

\usepackage{url}

\usepackage{wrapfig}
\usepackage{xspace}

\newcommand{\sysname}{MELODI\xspace}

\newcommand{\response}[1]{#1}

\usepackage{listings}

\colorlet{punct}{red!60!black}
\definecolor{background}{HTML}{EEEEEE}
\definecolor{delim}{RGB}{20,105,176}
\colorlet{numb}{magenta!60!black}

\def\BibTeX{{\rm B\kern-.05em{\sc i\kern-.025em b}\kern-.08em
    T\kern-.1667em\lower.7ex\hbox{E}\kern-.125emX}}
    
\begin{document}

\title{The Price of Prompting: Profiling Energy Use in Large Language Models Inference}

\author{Erik Johannes Husom, 
        Arda~Goknil,
        Lwin Khin Shar, and Sagar Sen 
\IEEEcompsocitemizethanks{
\IEEEcompsocthanksitem E.J. Husom, A. Goknil, and S. Sen are affiliated with SINTEF, Norway. L.K. Shar is affiliated with Singapore Management University, Singapore.\protect \\
 E-mails: erik.johannes.husom@sintef.no arda.goknil@sintef.no lkshar@smu.edu.sg sagar.sen@sintef.no 
}

\thanks{Manuscript received \response{June 2025; revised January 2026}.}}

\maketitle

\begin{abstract}

In the realm of Artificial Intelligence (AI), the deployment of Large Language Models (LLMs) has become fundamental to advancing technology and enriching user experiences across various applications. However, as these models grow in complexity and size, they pose significant computational and environmental challenges due to their substantial energy consumption. The increasing deployment of LLMs accentuates the need for sustainable practices to manage their energy demands effectively, especially in resource-constrained environments. In this paper, we introduce \sysname – \textbf{M}onitoring \textbf{E}nergy \textbf{L}evels and \textbf{O}ptimization for \textbf{D}ata-driven \textbf{I}nference – a multifaceted framework crafted to monitor and analyze the energy consumed during LLM inference processes. 
\response{Unlike existing tools that measure energy at the system level, \sysname uniquely combines process-level CPU monitoring via Scaphandre with GPU tracking via nvidia-smi, employing configurable temporal buffers to capture complete inference cycles with high precision.} \sysname enables detailed observations of power consumption dynamics and facilitates the creation of datasets reflective of energy efficiency across various deployment scenarios. We apply \sysname to profile inference-time energy consumption under a diverse set of scenarios. 
\response{Our findings reveal pronounced energy efficiency disparities: large models ($\geq$70B parameters) consume roughly two orders of magnitude more energy per token than smaller models, while response token length is a strong predictor of energy use (R$^2$ $\textgreater$ 0.95 in most settings). We further derive a predictive model with an R$^2$ value of 0.9962, utilizing response length, model type, and hardware. Prompt characteristics exhibit minimal correlation with energy consumption. In contrast, hardware choice has a substantial impact—laptop deployments consistently consume more energy than workstations for comparable models—highlighting significant opportunities for optimization and more sustainable LLM deployment.} 


\end{abstract}

\maketitle

\section{Introduction}
\label{sec:introduction}

The swift progression of artificial intelligence (AI) has precipitated the emergence of advanced natural language processing (NLP) systems. Large language models (LLMs) have ascended to prominence among these systems, proving indispensable across a vast spectrum of applications. Their utility ranges from simple text predictions to managing complex dialogues. In various domains, LLMs have shown great promise in automating and enhancing a variety of tasks that traditionally require significant human expertise and effort~\cite{vaswani2017attention, bommasani2021opportunities, brown2020language,bender2021dangers, radford2019language}. For instance, LLMs can assist in content generation~\cite{sudhakaran2024mariogpt, liang2024controllable, li2024pre}, customer support~\cite{wulf2024exploring, kolasani2023optimizing, chen2024large}, machine translation~\cite{li2024eliciting, zhang2023prompting, wang2023document}, sentiment analysis~\cite{zhong2023can, zhu2024model, zhang2023sentiment}, and legal document review, thereby streamlining operations across diverse sectors. The ability of LLMs to generate human-like text with high accuracy and relevance has made them invaluable tools for professionals, enabling them to focus on more complex and creative aspects of their disciplines.

\textbf{Context and Motivation.} As these models grow in computational complexity, their energy demands escalate correspondingly~\cite{rillig2023risks, weidinger2022taxonomy, weidinger2021ethical}. \response{The widespread deployment of LLMs across various applications, from code generation tools used in iterative workflows to general-purpose chatbots with hundreds of millions of users, makes individual usage accumulate to massive resource consumption. Unlike training, which occurs once per model, inference is a continuous operational cost that scales with usage.} In an epoch that prioritizes sustainable computing, the energy expenditure of LLMs is a consideration of mounting importance and one that commands focused attention~\cite{bommasani2021opportunities, liang2022holistic}. Therefore, the integration of LLMs in diverse applications from healthcare to customer service without compounding the carbon footprint becomes paramount, thus underscoring the need for techniques that can mitigate the energy demands of LLMs while maintaining their functionality and accessibility.


\textbf{Problem Statement.} Despite the widespread adoption of LLMs, 
understanding their energy implications remains a significant challenge~\cite{husom2025sustainable}. 
Prior studies have predominantly concentrated on the training phase of traditional machine learning (ML) models, with the inference phase often relegated to a secondary focus despite its cumulative energy footprint over the lifespan of a model~\cite{strubell2019energy}. Tools such as Green Algorithms~\cite{lannelongue2021green} provide estimations of carbon emissions for computational tasks and yet lack real-time monitoring capabilities. Furthermore, while initiatives like CodeCarbon~\cite{codecarbon} and its integration to ML pipelines~\cite{husom2024cain} offer a more automated tracking of emissions, they neither cater to the nuanced demands of LLMs nor provide granular data concerning inference-related energy consumption.

Recent research~\cite{liang2022holistic, samsi2023words,wilkins2024offline,wilkins2024hybrid, chien2023reducing,stojkovic2024towards,everman2023evaluating, argerich2024measuring} highlights the significant energy and carbon footprints of LLM configurations. Wilkins et al.~\cite{wilkins2024offline} introduce an offline, workload-based framework to optimize energy consumption for LLM inference on mixed GPU-CPU systems, focusing on energy usage and runtime based on token interactions. Their method 
faces limitations like computational overhead due to frequent sampling and lack of granularity. 
Conversely, Samsi et al.~\cite{samsi2023words} examine the energy consumption of GPUs in LLaMA models, excluding CPU impacts. \response{Prior work neither systematically applies a comprehensive framework for monitoring and analyzing LLM inference energy across diverse operational conditions, nor investigates real-time optimization across heterogeneous hardware. To address this gap, we aim to develop a holistic energy monitoring framework tailored to LLM deployments and use it to analyze how prompt characteristics and hardware configurations influence energy consumption. The framework should advance understanding of LLM energy use while enabling data-driven optimization strategies to support more efficient and sustainable AI engineering.}

\textbf{Our Design.} This paper introduces the \sysname framework designed to monitor and analyze energy usage during LLM inference. The urgent need for such a framework stems from the growing environmental and economic costs associated with the operation of LLMs. \sysname stands at the intersection of ML, software engineering, and sustainability, encapsulating a holistic approach. 
\sysname leverages two specialized power usage monitoring tools (Scaphandre~\cite{scaphandre} and Nvidia-smi~\cite{nvidia}). Scaphandre tracks the CPU’s power consumption for the LLM process, while nvidia-smi measures the GPU’s total power usage. For accurate measurements, the GPU must be exclusively used for the LLM inference, with no concurrent processes. \sysname records the energy consumption for each inference process, in addition to the prompt and the LLM’s response. Furthermore, it facilitates the collection of metadata associated with LLM inference tasks, enabling a deeper analysis of the relationship between task complexity and energy utilization.

\response{\textbf{Exploratory Study.} 
We conducted a study using \sysname to profile inference-time energy consumption under a diverse set of LLM inference scenarios. We deployed multiple open-source LLMs across heterogeneous hardware platforms, ranging from CPU-only laptops to GPU-equipped workstations and servers, and evaluated them on two representative prompt datasets (Alpaca~\cite{alpaca} and Code-Feedback~\cite{codefeedback}). For each inference request, \sysname recorded prompt–response pairs, token-level metadata, timestamps, and 
CPU/GPU power traces, which were integrated into an energy consumption dataset enabling comparative analyses across model types, model sizes, hardware configurations, and workload characteristics.}

\response{Our study reveals pronounced variability in LLM inference energy across deployment conditions: 70B-class models consume up to two orders of magnitude more energy per token than smaller models, and laptop deployments are generally less efficient than workstations, with measurable differences even between models of similar size. Energy consumption is driven primarily by response characteristics (response length and duration), while prompt complexity features show weak associations. Using response length, we achieve highly accurate energy prediction and derive an interpretable model with near-perfect performance ($R^2 = 0.9962$) based on response length, model type, and hardware. Finally, repeated runs show model-dependent variability, and comparisons with other monitoring tools expose substantial measurement discrepancies, highlighting the need for robust inference-time energy monitoring.}

To summarize, our contributions include:



\begin{itemize}

    \item \textbf{\sysname Framework:} We introduce \sysname, an open-source and extensible framework for fine-grained monitoring and analysis of CPU and GPU energy consumption during LLM inference, enabling reproducible energy profiling at the level of individual inference requests.
    
    

    \item \textbf{Energy Consumption Dataset:} Using \sysname, we construct and release a comprehensive inference-time energy dataset spanning multiple hardwares, LLM families and model sizes, and prompt datasets, supporting systematic benchmarking of energy efficiency across deployment conditions.



    \item \textbf{Empirical Characterization of Energy Drivers:} We conduct an extensive exploratory study identifying how model size, model type, hardware configuration, and prompt/response characteristics influence inference energy consumption, providing evidence-based insights for energy-aware deployment.

    \item \response{\textbf{Prediction and Interpretable Modeling:} We develop and assess predictive models for inference energy, 
    demonstrating that energy can be reliably modeled from response length, model type, and hardware.}
    
\end{itemize}

The paper is structured as follows. Section~\ref{sec:background} presents the background regarding 
energy monitoring tools. In Section~\ref{sec:dataset_collection}, we present \sysname. Section~\ref{sec:experiments} reports on the experiments. Section~\ref{sec:discussion} presents some future work. Section~\ref{sec:related_work} discusses the related work. We conclude the paper in Section~\ref{sec:conclusion}.

\section{Background: Monitoring Technologies}
\label{sec:background}

Advanced monitoring tools and technologies are essential for tracking and optimizing energy consumption. 
In this context, \sysname employs the following tools:  

\textbf{NVIDIA System Management Interface (nvidia-smi)}~\cite{nvidia} is a command-line utility, based on NVIDIA Management Library (NVML)~\cite{nvidia-nvml}, providing the real-time monitoring of GPU, including power consumption, for systems equipped with NVIDIA GPUs. This allows administrators to query GPU device state and with the appropriate privileges, permits administrators to modify GPU device state~\cite{nvidia}. 


\textbf{Scaphandre}~\cite{scaphandre} is a metrology agent designed to enhance energy management across various computing environments. It is one of the few tools providing detailed insights into the energy usage of computing systems at the process level~\cite{jay2023experimental}. It offers comprehensive features including the ability to measure power consumption directly on bare metal hosts and within qemu/kvm virtual machines from the host. Additionally, Scaphandre facilitates the integration of power metrics into virtual machines as if they were bare metal setups and supports exporting these metrics as a Prometheus HTTP exporter~\cite{prometheus-exporter}. 






\section{\sysname for Monitoring Energy Usage}
\label{sec:dataset_collection}


\begin{figure}
    \centering
    \includegraphics[width=0.80\linewidth]{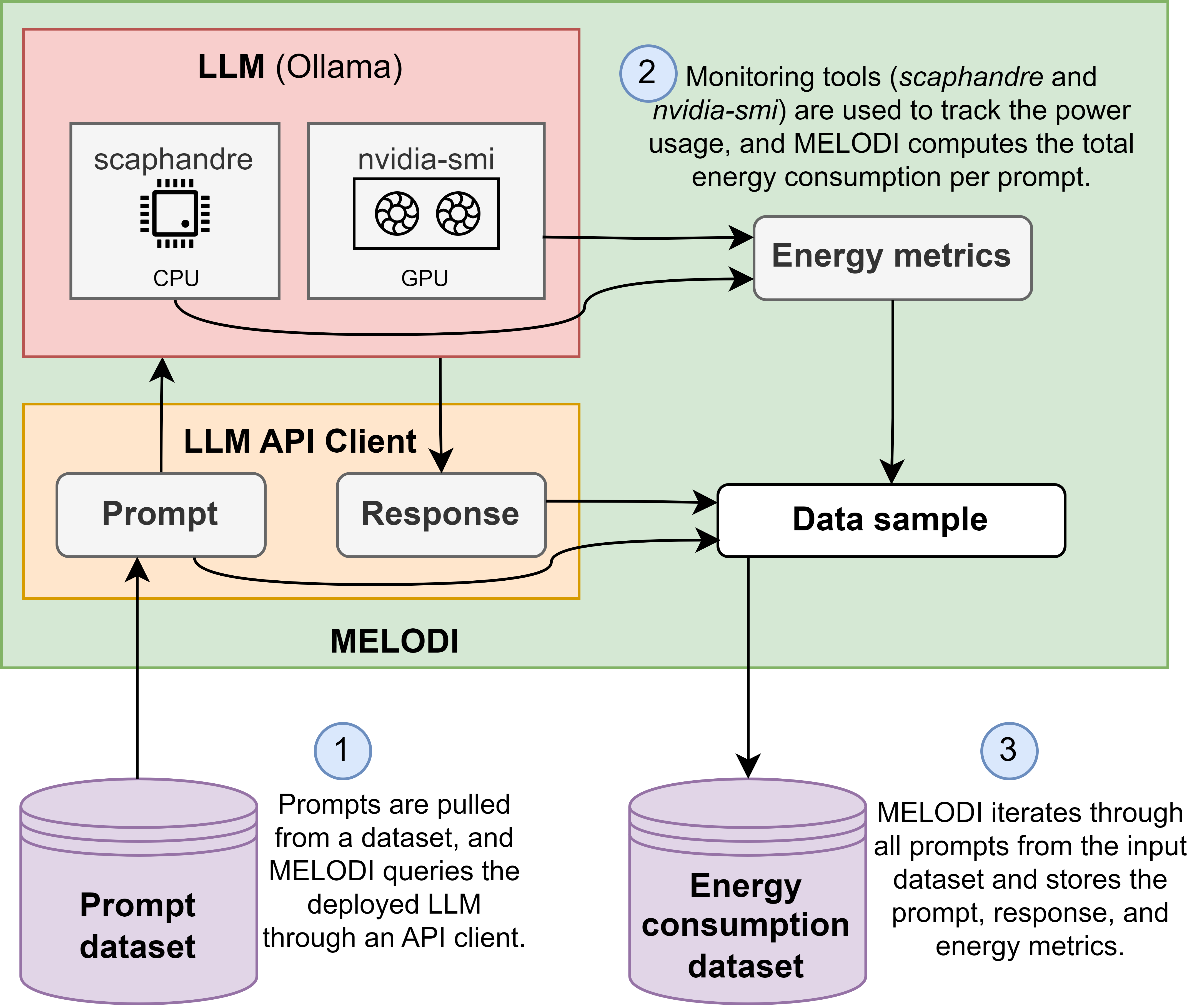}
    \vspace*{-0.20em}
    \caption{Overview of the \sysname framework.}
    \label{fig:framework}
    \vspace*{-1.7em}
\end{figure}

To analyze LLM energy consumption, we developed the \sysname framework, which monitors energy use during LLM inference. Figure~\ref{fig:framework} outlines its architecture, highlighting its process of sampling prompts, submitting them to an LLM, and recording the prompts, responses, metadata, and energy metrics. This collected data allows for a comprehensive analysis of energy trends during LLM operations.

\begin{itemize}

   \item \textbf{Input Data (Prompt Dataset in Figure~\ref{fig:framework}):} \sysname extracts prompts from a pre-existing dataset 
   to query a deployed LLM through an API client.
   

   \item \textbf{Power Usage Monitoring and Data Collection:} \sysname utilizes two complementary power monitoring tools (i.e., Scaphandre~\cite{scaphandre} and nvidia-smi~\cite{nvidia}) to track the power consumption of LLMs like the Ollama model in Figure~\ref{fig:framework}. Scaphandre records CPU power consumption specific to the LLM process, while nvidia-smi measures total GPU power usage, requiring exclusive GPU operation (no other processes on the GPU) during measurements for accuracy. \sysname calculates energy consumption for each prompt, aggregating data on the prompt, response, and energy metrics for detailed analysis.



   \item \textbf{Dataset Generation:} \sysname iterates through all prompts from the prompt dataset. 
   The data obtained for each prompt are aggregated into a dataset, encapsulating the energy consumption of LLM inference.

\end{itemize}

\sysname relies on LLM services that can be queried via API endpoints for generating responses. Our implementation supports services adhering to Ollama or OpenAI API specifications, covering most LLM deployment services that provide natural language responses and metadata such as timestamps and token lengths for prompts and responses. The landscape of LLM deployment tools is diverse, featuring platforms such as Llama.cpp~\cite{llama.cpp}, GPT4All~\cite{anand2023gpt4all}, vLLM~\cite{kwon2023efficient}, Llamafile~\cite{llamafile}, and Ollama~\cite{ollama}. We utilized Ollama for our experiments due to its easy setup, compatibility with both CPU and GPU architectures, and support for several open-source LLMs.

Scaphandre operates with process-level granularity, monitoring and recording the power usage of active processes on a computing device. We apply a regular expression to the process names to isolate the specific processes associated with the LLM service. The frequency at which Scaphandre samples power metrics is configurable with granularity fine enough to reach the nanosecond scale. Although a higher sampling frequency increases the resolution and accuracy of the data, it also requires more storage for data retention and increases the energy consumption of Scaphandre. Therefore, we also monitor the energy consumption of Scaphandre, enabling an assessment of the energy overhead incurred by the measurement process.

Nvidia-smi tracks the power usage of the entire GPU and cannot distinguish among multiple processes concurrently utilizing the GPU. To guarantee precision, we limit the GPU to solely operating the LLM inference process. The two tools measure the power consumption $P$ of a process in microwatts ($\mu$W). We compute the energy consumption $E$ by integrating the power consumption over the duration $t$, applying the trapezoidal rule. With $t$ (in seconds), we convert $E$ into kilowatt-hours (kWh) by dividing by $3.6 \times 10^{12}$, facilitating a standardized energy usage assessment.

\begin{wrapfigure}{r}{.27\textwidth}
    \vspace*{-0.8em}

    \begin{minipage}{\linewidth}
    \centering
    \includegraphics[width=\linewidth]{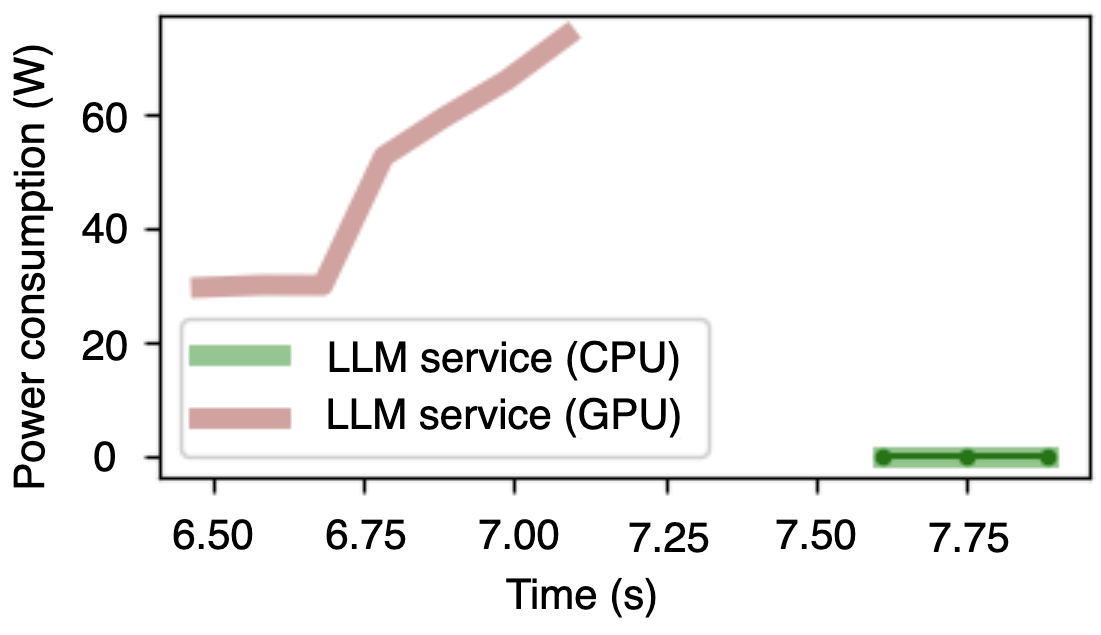}
    \subcaption{Power draw during inference ($M_0 = M_1 = 0$ and $R_0 = R_1 = 0$).}
    \label{fig:power1}\par\vfill
    \includegraphics[width=\linewidth]{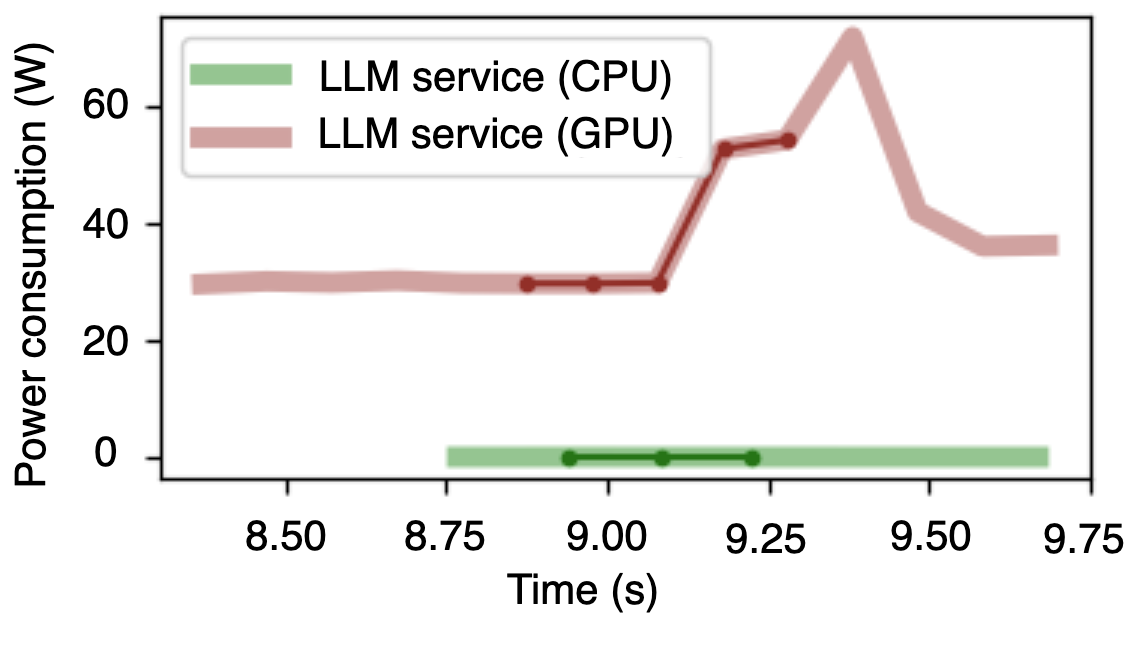}
    \subcaption{Power draw during inference ($M_0 = M_1 = 0.5$ seconds and $R_0 = R_1 = 0$).}
    \label{fig:power2}
    \includegraphics[width=\linewidth]{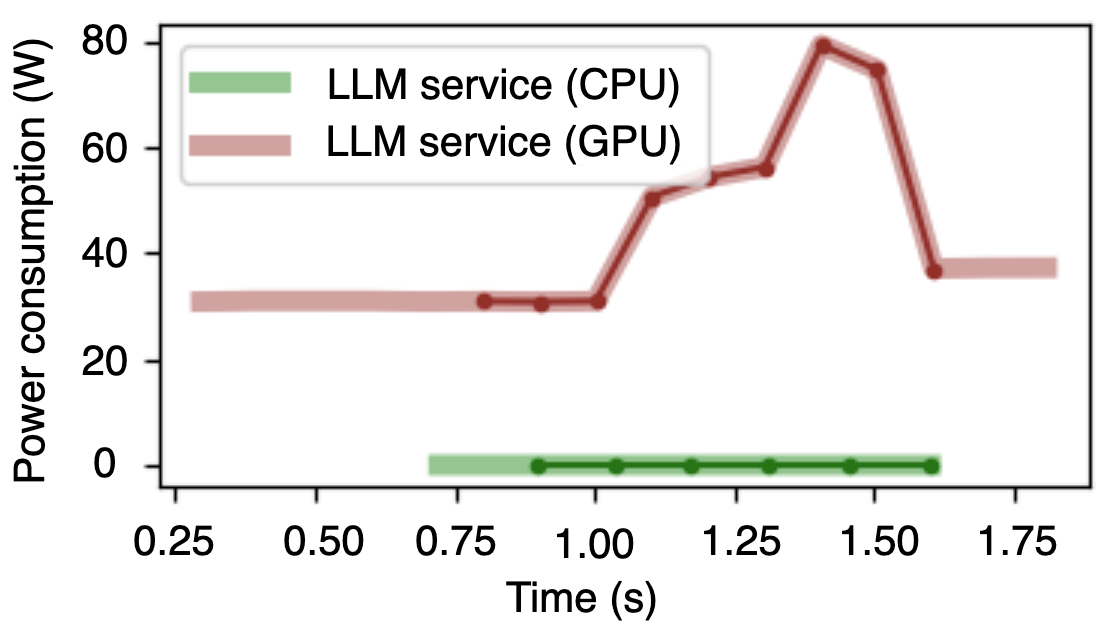}
    \subcaption{Power draw during inference ($M_0 = M_1 = 0.5$ seconds, and $R_0 = 0$ and $R_1 = 0.2$ seconds).}
    \label{fig:power3}
\end{minipage}
\vspace*{-0.3em}
\caption{Power draw during a prompt inference as monitored by \sysname with varying buffer settings. Thick semi-transparent lines indicate active monitoring intervals by Scaphandre (CPU) and nvidia-smi (GPU), while the solid line with dots are data points recorded by \sysname.}\label{fig:power_profiles}
\vspace*{-1.4em}
\end{wrapfigure}


In each inference, \sysname activates Scaphandre to monitor CPU power usage at the process level, targeting the LLM service to ensure measurement accuracy. Concurrently, nvidia-smi tracks GPU power usage at the device level, necessitating the restriction of GPU activities solely to LLM inference to eliminate external application interference. \sysname logs the prompt, response, token counts, and the timestamps marking the start and end of the inference, while recording the energy consumption and power draw as a time series. 

To enhance energy measurement accuracy during inference, we implement two types of buffers in \sysname to compensate for monitoring tool delays. The first, \textit{monitoring buffer} ($M$), introduces a delay before and after the inference process to ensure Scaphandre and nvidia-smi are fully operational also before and after the inference. This buffer is divided into two configurable intervals: $M_0$, the delay from monitoring startup to inference start, and $M_1$, the delay from inference end to monitoring shutdown, effectively preventing data loss at the critical start and end inference points due to tool response times.


Figures~\ref{fig:power1} and \ref{fig:power2} demonstrate the impact of employing a monitoring buffer on data capture. Figure~\ref{fig:power1}, utilizing no buffer ($M_0 = M_1 = 0$), shows that nvidia-smi fails to capture data during extremely brief inference runs, resulting in incomplete energy measurements. Conversely, Figure~\ref{fig:power2}, using a 0.5-second buffer for pre- and post-inference ($M_0 = M_1 = 0.5$), shows improved reliability in data capture, accounting even for brief inference periods. \response{We tested multiple values of $M$ (Table~\ref{tab:m_buffer}) and found that $M=0.5$ provides optimal data capture without including unnecessary measurements. A monitoring interval was considered successful when the power profile was stable both before and after the inference window.}



Even with a monitoring buffer, we observe that GPU power spikes occasionally persist briefly post-inference, as depicted in Figure~\ref{fig:power2}. To address this, we introduce a \textit{recording buffer} ($R$), extending the monitoring period before and/or after the actual inference window. This buffer is split into $R_0$, extending the recording time before inference, and $R_1$, extending after. $R_1$ is crucial for capturing lingering GPU power spikes after inference concludes, ensuring comprehensive data collection (see Figure~\ref{fig:power3}). This delayed power consumption decline, also observed in other studies~\cite{yang2024parttime}, underscores the necessity of an end buffer for precise measurements.

\response{Determining optimal buffer values of $R$ involves a trade-off between measurement completeness and estimation bias. Larger buffers ensure full capture of transient power spikes but can overestimate energy by including baseline consumption outside the inference window. Smaller buffers reduce this bias but may miss lingering power decay, particularly on GPUs (Figure~\ref{fig:power3}). As shown in Table \ref{tab:r0_buffer}, $R_0 = 0.1$ achieves complete capture but consistently includes 0.1 seconds of pre-inference baseline power, whereas $R_0 = 0.0$ removes this offset while still capturing 94\% of cases. Similarly, Table~\ref{tab:r1_buffer} shows that $R_1 = 0.3$ guarantees full capture of post-inference power decay but often extends beyond the actual spike, while $R_1 = 0.2$ provides a better balance, achieving 88\% completeness with reduced baseline inclusion. We therefore set $R_0 = 0.0$ and $R_1=0.2$ for all experiments, balancing measurement accuracy and completeness. This choice avoids systematic overestimation while tolerating a small, controlled loss of data.}


\begin{table}[t]
    \centering
    \caption{\response{Capture completeness rate (CCR) for different buffer settings, measured as the percentage of runs (50 per setting) in which the full power spike was captured. In some cases, achieving 100\% CCR includes excess baseline power outside the inference window.}}
    \label{tab:monitoring_buffer}
    \vspace*{-0.8em}
    \begin{subtable}[t]{0.32\columnwidth}
        \centering
        \caption{$M = M_0 = M_1$}
        \label{tab:m_buffer}
        \begin{tabular}{lr}
        \toprule
        $M$ (s)& CCR \\
        \midrule
        0.2 & 0\% \\
        0.4 & 48\% \\
        0.5 & 100\% \\
        1.0 & 100\% \\
        \bottomrule
        \end{tabular}
    \end{subtable}
    \hfill
    \begin{subtable}[t]{0.32\columnwidth}
        \centering
        \caption{$R_0$}
        \label{tab:r0_buffer}
        \begin{tabular}{lr}
        \toprule
        $R_0$ (s)& CCR \\
        \midrule
        0.0 & 94\% \\
        0.1 & 100\% \\
        0.2 & 100\% \\
        0.3 & 100\% \\
        \bottomrule
        \end{tabular}
    \end{subtable}
    \hfill
    \begin{subtable}[t]{0.32\columnwidth}
        \centering
        \caption{$R_1$}
        \label{tab:r1_buffer}
        \begin{tabular}{lr}
        \toprule
        $R_1$ (s)& CCR \\
        \midrule
        0.0 & 34\% \\
        0.1 & 36\% \\
        0.2 & 88\% \\
        0.3 & 100\% \\
        \bottomrule
        \end{tabular}
    \end{subtable}
    \vspace*{-1.4em}
\end{table}

\section{Experiments}
\label{sec:experiments}

We analyze the energy consumption dataset to address three Research Questions (RQ)s:

\begin{itemize}

\item \textit{\textbf{RQ1. How does the energy consumption of LLM inference vary across different hardware, models, and prompt datasets?}} 

\item \textit{\textbf{RQ2. What is the relationship between prompt complexity, response characteristics, and the energy consumption of LLMs during the inference process?}} 

\item \textit{\textbf{RQ3. Can we develop a predictive model accurately forecasting the energy consumption of LLMs based on prompt features and response characteristics?}} 


\item \textit{\textbf{RQ4. To what extent can LLM inference energy consumption be modeled using an interpretable mathematical function of response length, model type, and hardware configuration?}}

\item \textit{\textbf{RQ5. How does the variability in energy consumption for the same prompt manifest across multiple executions and different models?}}





\item \textit{\textbf{RQ6. What are the differences in energy measurements among \sysname and other tools, and what factors contribute to these variations?}} 

\end{itemize}



\subsection{Subjects of the Experiments}

This section outlines the energy consumption dataset collected with \sysname, using diverse computing machines (from high-end servers to laptops) and multiple prompt datasets for several LLMs. 
Table~\ref{tab:promptdatasets} lists the prompt datasets in dataset collection, i.e., \texttt{Alpaca}~\cite{alpaca} and \texttt{Code-Feedback}~\cite{codefeedback}, and their average energy consumption and response token length. 
The \texttt{Alpaca} dataset, produced using OpenAI's text-davinci-003 engine, contains 52,000 prompts designed to enhance instruction-following capabilities in language models. 
The dataset primarily includes text generation tasks and is tailored to train pre-trained language models to follow complex instructions more effectively. The \texttt{Code-Feedback} dataset supports the OpenCodeInterpreter model~\cite{opencodeinterpreter} designed to refine code by integrating code execution and human feedback. This dataset features 68,000 multi-turn interactions, combining user instructions with compiler responses to enhance model training in coding scenarios. 


\begin{table}[h]
    \vspace*{-0.80em}
    \scriptsize
    \caption{Prompt datasets.}
    \label{tab:promptdatasets}
    \vspace*{-0.80em}

    \centering
    \begin{tabular}{crr}
    \hline
    \begin{tabular}{@{}c@{}}Prompt \\ dataset\end{tabular} & \begin{tabular}{@{}c@{}}Average energy \\ consumption (kWh)\end{tabular} & \begin{tabular}{@{}c@{}} Average response \\ token length \end{tabular} \\
    \hline
    Alpaca & 4.49e-04 & 213.98 \\  
    Code-feedback & 5.68e-04 & 405.96 \\                                                      
    \hline
    \end{tabular}
    \vspace*{-0.90em}
\end{table}

\begin{table}[h]
    \vspace*{-0.80em}
    \scriptsize
    \caption{LLMs in the experiments.}
    \label{tab:models}
    \vspace*{-0.80em}

    \centering
    \begin{tabular}{crr}
    \hline
    Model name/size & \begin{tabular}{@{}c@{}}Average energy \\ consumption (kWh)\end{tabular} & \begin{tabular}{@{}c@{}} Average response \\ token length\end{tabular} \\
    \hline
codellama-7b & 2.35e-04 & 535.03 \\
codellama-70b & 3.33e-03 & 342.64 \\  
gemma-2b & 8.18e-05 & 279.11 \\
gemma-7b & 1.36e-04 & 249.33 \\ 
llama3-8b & 1.34e-04 & 255.20 \\  
llama3-70b & 2.26e-03 & 251.46 \\
phi-2b & 3.14e-05 & 179.30 \\
qwen2-7b & 8.37e-05 & 199.34 \\
qwen2-72b & 2.25e-03 & 210.82 \\
    \hline
    \end{tabular}
    \vspace*{-0.90em}

\end{table}

Table~\ref{tab:models} lists the LLMs with their average energy consumption and response token lengths. The models range from \texttt{codelama-7b} to \texttt{llama3-70b}, with energy consumption varying from as low as \(8.20 \times 10^{-5}\) kWh for \texttt{gemma-2b} to \(3.33 \times 10^{-3}\) kWh for \texttt{codellama-70b}. The average response token lengths vary, with \texttt{codelama-7b} generating the longest responses at 535.03 tokens, while \texttt{phi-2b} records shorter lengths at 179.30 tokens. 

\begin{table*}[h]
    \scriptsize
    \caption{Hardware used in the data collection.}
    \label{tab:hardware}
    \vspace*{-1.10em}

    \centering
    \begin{tabular}{ccccc}
    \hline
    Machine & CPU & Memory & GPU & GPU memory \\
    \hline
    Server      & AMD EPYC 7643 48-Core Processor & 528GB & NVIDIA RTX A5000 & 24GB \\
    Workstation & Intel Xeon W-2223 8-core @ 3.6GHz & 128GB & NVIDIA RTX A2000 & 12GB \\
    Laptop 1    & Intel i5 11th Gen 12-core @ 2.4GHz & 16GB & None & None \\
    Laptop 2    & Intel i7 10th Gen 12-core @ 2.7GHz & 32GB & NVIDIA Quadro RTX 4000 & 8GB \\
    \hline
    \end{tabular}
\end{table*}

\begin{table*}[h]
    \scriptsize
    \centering
    \caption{Overview of the energy consumption dataset.}
    \label{tab:datasets}
    \vspace*{-1.10em}
    \begin{tabular}{cccccrrr}
    \hline
         ID & Prompt dataset & Model & Model size & Hardware & No. of prompts & \begin{tabular}{@{}c@{}}Average energy consumption \\ per response (kWh)\end{tabular} & \begin{tabular}{@{}c@{}}Average response \\ token length\end{tabular}\\
     \hline
1 & code-feedback & codellama & 7b & laptop1 & 5347 & 1.85e-04 & 403.63 \\
2 & code-feedback & codellama & 7b & laptop2 & 3572 & 2.46e-04 & 518.40 \\
3 & code-feedback & codellama & 7b & workstation & 1000 & 2.73e-04 & 685.06 \\
4 & code-feedback & codellama & 70b & workstation & 1000 & 3.33e-03 & 342.64 \\
5 & alpaca & gemma & 2b & laptop1 & 5347 & 1.85e-04 & 401.64 \\
6 & alpaca & gemma & 2b & laptop2 & 5101 & 4.70e-05 & 181.52 \\
7 & alpaca & gemma & 2b & workstation & 1000 & 3.84e-05 & 184.30 \\
8 & code-feedback & gemma & 2b & laptop2 & 5008 & 7.33e-05 & 303.98 \\
9 & code-feedback & gemma & 2b & workstation & 1000 & 6.60e-05 & 324.14 \\
10 & alpaca & gemma & 7b & laptop2 & 5099 & 9.81e-05 & 160.60 \\
11 & alpaca & gemma & 7b & workstation & 1000 & 8.05e-05 & 164.90 \\
12 & code-feedback & gemma & 7b & laptop2 & 3405 & 2.00e-04 & 332.62 \\
13 & code-feedback & gemma & 7b & workstation & 1000 & 1.65e-04 & 339.21 \\
14 & alpaca & llama3 & 8b & laptop2 & 5101 & 1.34e-04 & 255.20 \\
15 & alpaca & llama3 & 70b & server & 1026 & 2.26e-03 & 251.46 \\
16 & alpaca & phi & 2b & laptop2 & 674 & 3.22e-05 & 221.63 \\
17 & alpaca & phi & 2b & workstation & 1000 & 3.15e-05 & 136.98 \\
18 & alpaca & qwen2 & 7b & laptop2 & 821 & 1.22e-04 & 300.70 \\
19 & alpaca & qwen2 & 7b & workstation & 1000 & 1.11e-04 & 97.99 \\
20 & alpaca & qwen2 & 72b & workstation & 1000 & 2.25e-03 & 210.82 \\
     \hline
    \end{tabular}
    \vspace*{-1.4em}
\end{table*}

Table~\ref{tab:hardware} details the hardware used in our data collection. The diversity of the hardware covers a wide range of processing power, memory capacities, and graphics capabilities, crucial for a comprehensive analysis of energy consumption across different computing environments. 

Table~\ref{tab:promptdatasets} summarizes the energy dataset collected by MELODI across multiple hardware setups and LLMs under diverse operational conditions. Each sample includes the prompt and response, prompt/response token counts, API timestamp, time-series power traces for the LLM process and Scaphandre during monitoring, and the aggregated inference energy consumption.

\subsection{Experiments Setup}

\emph{\textbf{Data Preparation.}} Our raw data consists of power usage over time, which we aggregate to calculate the energy consumption for each inference process as outlined in Section~\ref{sec:dataset_collection}. These aggregated values are compiled into distinct energy consumption datasets, categorized by the specific hardware, model, and prompt dataset utilized during the inference.



\emph{\textbf{Statistical Analysis.}} Our analysis of the energy consumption datasets focuses on two metrics: energy per token and energy per response. Energy consumption per token is crucial as it evaluates resource use across different inference setups regardless of text volume. Meanwhile, energy per response is beneficial for estimating resource usage relative to the input prompt. We conduct statistical analyses on these metrics to compare across model types, sizes, hardware setups, and the specific prompt datasets used during inference, providing a comprehensive view of energy dynamics.


\emph{\textbf{Visualization.}} We employ box plots to visualize the distribution of energy consumption for each model and hardware. 
The box plots depict the median energy consumption as a red line, the interquartile range (IQR) with the main box, and extend to 1.5 times the IQR with whiskers, providing a clear summary of variability and central tendency.

\emph{\textbf{Interpretation.}} We interpret results through a comparative analysis of energy consumption across different scenarios. The relationship between prompt complexity and energy consumption is assessed using the Pearson correlation coefficient, which ranges from 0 to 1 to indicate variable correlation. Additionally, predictive models for energy consumption are evaluated using the $R^2$ score.

\subsection{Results}

\subsubsection{RQ1 (energy consumption of LLM inference across different setups)}



\begin{figure*}
    \centering
    \begin{subfigure}{0.81\linewidth}
        \includegraphics[width=0.86\linewidth]{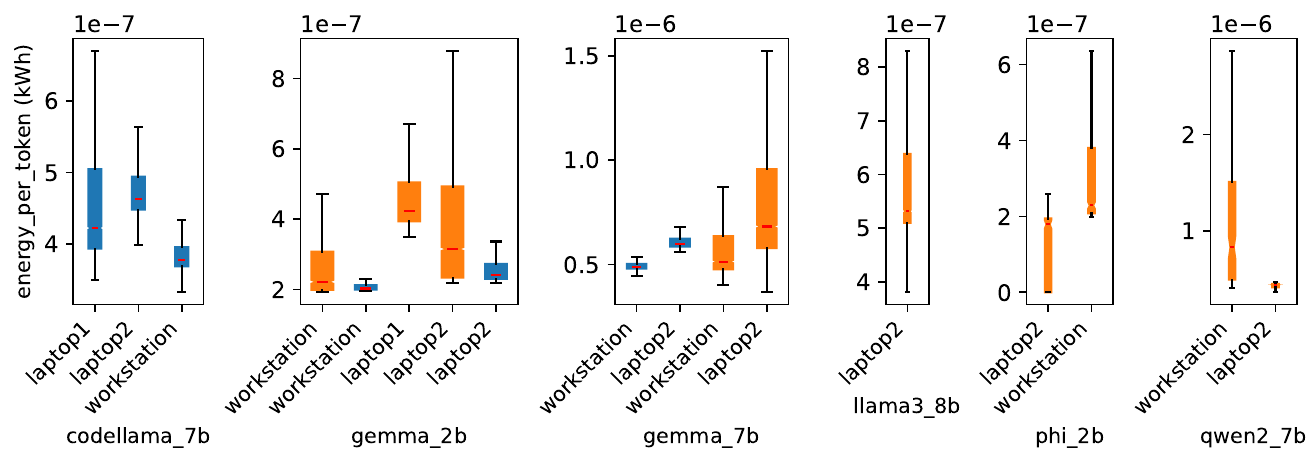}
        \caption{Comparison of energy consumption per token across models with 2-8 billion parameters.}
        \label{fig:model_comparison}
    \end{subfigure}
      \begin{subfigure}{0.81\linewidth}
        \includegraphics[width=0.86\linewidth]{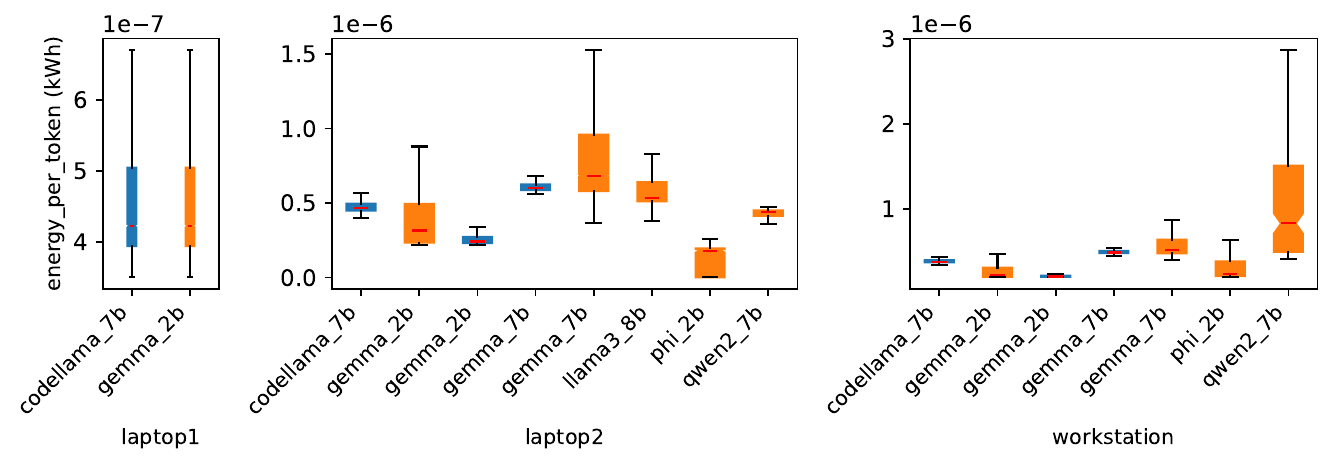}
        \caption{Comparison of energy consumption per token across hardware setups (Codellama-70b is excluded for the workstation to not skew the plot).}
        \label{fig:hardware_comparison}
    \end{subfigure}
    \caption{Comparative energy consumption per token across inference setups.}
    \label{fig:energy_per_token_comparison}
    \vspace*{-1.1em}
\end{figure*}

\begin{figure}[t]
    \centering
    \vspace*{-1.0em}
    \includegraphics[width=0.70\linewidth]{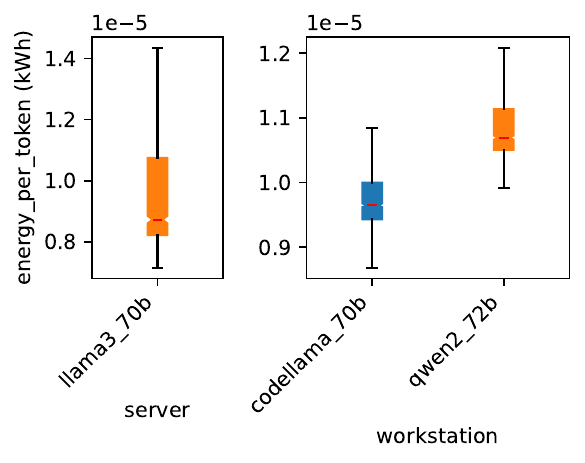}
    \vspace*{-0.8em}
    \caption{Comparison of energy consumption per token across models with 70-72 billion parameters.}
    \label{fig:model_comparison_large}
    \vspace*{-1.8em}

\end{figure}

To address RQ1, we performed a comparative analysis of energy consumption across various LLMs operating on different hardware (see Figures~\ref{fig:energy_per_token_comparison} and \ref{fig:model_comparison_large}). 

\emph{\textbf{Model Size.}} Figures~\ref{fig:model_comparison} and \ref{fig:model_comparison_large} categorize energy consumption per token by the model utilized for inference. Notably, the largest models (\texttt{codellama-70b} with \texttt{Code-Feedback}, \texttt{qwen2-72b} and \texttt{llama3-70b} with \texttt{Alpaca}) exhibit energy usage approximately 100 times greater than their smallest counterparts (\texttt{codellama-7b} with \texttt{Code-Feedback}, \texttt{qwen2-7b} and \texttt{llama3-8b} with \texttt{Alpaca}). Additionally, the larger models (\texttt{gemma-7b} with both \texttt{Alpaca} and \texttt{Code-Feedback}) show an energy consumption that is roughly ten times higher than their smaller versions (\texttt{gemma-2b} with \texttt{Alpaca} and \texttt{Code-Feedback}).


\emph{\textbf{Hardware.}} Figure~\ref{fig:hardware_comparison} organizes results by the hardware. 
Energy consumption is noticeably higher for the laptops than the workstation. Laptop 1 (CPU-only) shows similar energy usage when operating different models (\texttt{codellama-7b} with \texttt{Code-Feedback} and \texttt{gemma-2b} with \texttt{Alpaca}), suggesting inefficiencies in CPU-based processing for LLM tasks since we would expect a significant difference when running a 2b model vs a 7b model. Comparisons between the workstation and Laptop 2, despite the former's more robust GPU (12GB vs. 8GB), reveal roughly equivalent energy usage, underscoring potential disparities in hardware efficiency.


\emph{\textbf{Model.}} In energy assessments of different models of the same size (\texttt{codellama-7b} with \texttt{Code-Feedback} vs. \texttt{gemma-7b} with \texttt{Code-Feedback}) run on the workstation and Laptop 2, \texttt{gemma-7b} displayed higher energy consumption than \texttt{codellama-7b}. This observation suggests that \texttt{codellama-7b} is a slightly more energy-efficient model than \texttt{gemma-7b}. \texttt{phi-2b} shows comparable energy consumption per token to the \texttt{gemma-2b} of similar size but has a slightly higher consumption on the workstation.



\emph{\textbf{Prompt Datasets.}} To evaluate the energy consumption per output token across two prompt datasets, we analyzed the box plots for \texttt{gemma-2b} and \texttt{gemma-7b} in Figure~\ref{fig:model_comparison}, the only models tested with both prompt datasets. The analysis reveals that, except in one instance, the \texttt{Alpaca} dataset consistently led to higher energy consumption per token. 
The IQR, representing variability in energy consumption, is notably wider in all cases, suggesting greater fluctuation in energy use when using the \texttt{Alpaca} dataset.



\response{Table~\ref{tab:promptdatasets} facilitates a comparison of energy consumption per prompt, revealing that the median consumption for the \texttt{Code-Feedback} prompt dataset exceeds that of the \texttt{Alpaca} dataset during inference. This discrepancy likely arises from longer response length of \texttt{Code-Feedback} prompts. The response length of \texttt{Code-Feedback} prompts averages at 405.96 tokens whereas the response length of \texttt{Alpaca} prompts averages at 213.98 tokens.}

\begin{tcolorbox}[boxsep=2pt,left=2pt,right=2pt,top=2pt,bottom=2pt]
\textbf{RQ1 Conclusion.} 
Larger models like \texttt{codellama-70b} and \texttt{llama3-70b} use roughly 100 times more energy per token than smaller ones. Energy demands are higher for LLMs running on laptops than workstations, particularly due to inefficiencies in CPU processing. Models of the same size exhibit varying energy efficiencies. The \texttt{Code-Feedback} dataset leads to greater energy use per token than \texttt{Alpaca}, likely due to longer response lengths.


\end{tcolorbox}

%

\subsubsection{RQ2 (relationship between prompt complexity, response characteristics and the energy consumption)}


To respond to RQ2, we used Python libraries, namely, Spacy~\cite{spacy2}, nltk~\cite{bird2009natural}, textblob~\cite{loria2018textblob}, and textstat~\cite{textstat}, to derive \response{55} text-based features. We integrated the features with our energy consumption dataset and calculated the Pearson correlation coefficient between energy consumption per prompt and each feature. 


\response{We extracted prompt features spanning multiple categories, including lexical richness and word-level complexity (e.g., average word length, long/polysyllabic word counts, lexical diversity), part-of-speech frequencies (e.g., noun and adjective counts), readability indices (e.g., Flesch, SMOG, Coleman–Liau, Dale–Chall), sentence-level structure (e.g., mean sentence length, punctuation counts), and sentiment scores. Table~\ref{tab:correlations} reports the top 20 correlated features.}

Our analysis revealed that significant correlations with energy consumption per prompt are primarily linked to response characteristics rather than the prompt complexity. Key factors such as response duration, total duration, and response token length show strong positive correlations with energy consumption (with coefficients of 0.770, 0.766, and 0.755, respectively). This observation suggests that energy consumption escalates with an increase in response tokens due to more extensive processing within the model, highlighting that longer and more time-intensive responses drive higher energy usage.


\begin{table}[t]
    \scriptsize
    \caption{Correlations between text features and energy consumption per response across all samples. Features are derived from the prompt unless marked with (*), which indicates response-based features.}
    \vspace*{-0.8em}
    \label{tab:correlations}
    \centering
\begin{tabular}{lr|lr}
\hline
Feature & Correlation & Feature & Correlation \\
\hline
*response\_duration & 0.770 & reading\_time & 0.076 \\
*total\_duration & 0.766 & lexicon\_count & 0.075 \\
*response\_token\_length & 0.755 & adverb\_count & 0.073 \\
prompt\_duration & 0.129 & prompt\_token\_length & 0.073 \\
adj\_count & 0.101 & stop\_word\_count & 0.070 \\
polysyllabcount & 0.087 & noun\_count & 0.070 \\
long\_word\_count & 0.085 & verb\_count & 0.068 \\
syllable\_count & 0.079 & monosyllabcount & 0.067 \\
letter\_count & 0.079 & word\_count & 0.065 \\
char\_count & 0.076 & sentence\_count & 0.058 \\

\hline
\end{tabular}
\vspace*{-1.40em}

\end{table}

We observe only low positive correlations with prompt complexity features. Attributes like prompt duration, adjective count, and syllable count demonstrate modest correlations with energy consumption, with coefficients of 0.129, 0.101, and 0.087, respectively. These results indicate that prompt complexity (given our set of features) has a minimal impact on energy consumption. Instead, the length and duration of responses are more significant factors. This insight suggests that optimizing the response generation process could be more effective in reducing energy consumption than merely simplifying the input prompts.

\begin{tcolorbox}[boxsep=2pt,left=2pt,right=2pt,top=2pt,bottom=2pt]
\textbf{RQ2 Conclusion.} 
The results show that response characteristics such as token length and duration are strongly correlated with energy usage, indicating higher consumption with longer responses. Conversely, prompt complexity features like adjective and syllable counts exhibit low correlations, suggesting minimal impact on energy use. Managing response generation may offer more significant energy savings than simplifying prompts.

\end{tcolorbox}

\subsubsection{RQ3 (predictive model for LLM energy consumption)}

\response{To address RQ3, we trained ML models on the energy dataset using either prompt features or response characteristics as inputs. We evaluated Random Forest (RF), Gradient Boosting (GB), Linear Regression (LR), XGBoost (XGB), Decision Tree (DT), and Support Vector Machine (SVM), training two models per dataset: one using response token length and one using prompt-derived features.}

\begin{figure}[t]
    \centering
    \includegraphics[width=1.0\linewidth]{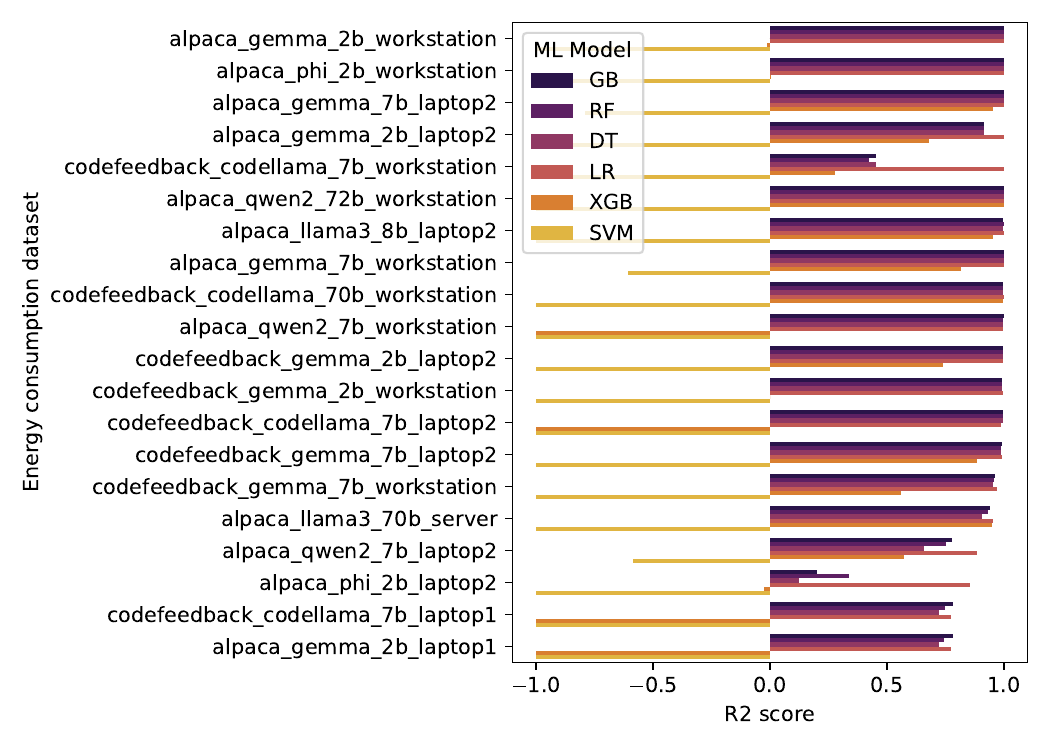}
    \vspace*{-1.9em}
    \caption{\response{R$^2$ scores across different ML algorithms, using only response length for forecasting energy consumption per response. Negative R$^2$ scores are clipped at -1.0.}}
    \label{fig:r2_ml_models_comparison_response}
    \vspace*{-0.8em}

\end{figure}

\begin{figure}[t]
    \centering
    \includegraphics[width=1.0\linewidth]{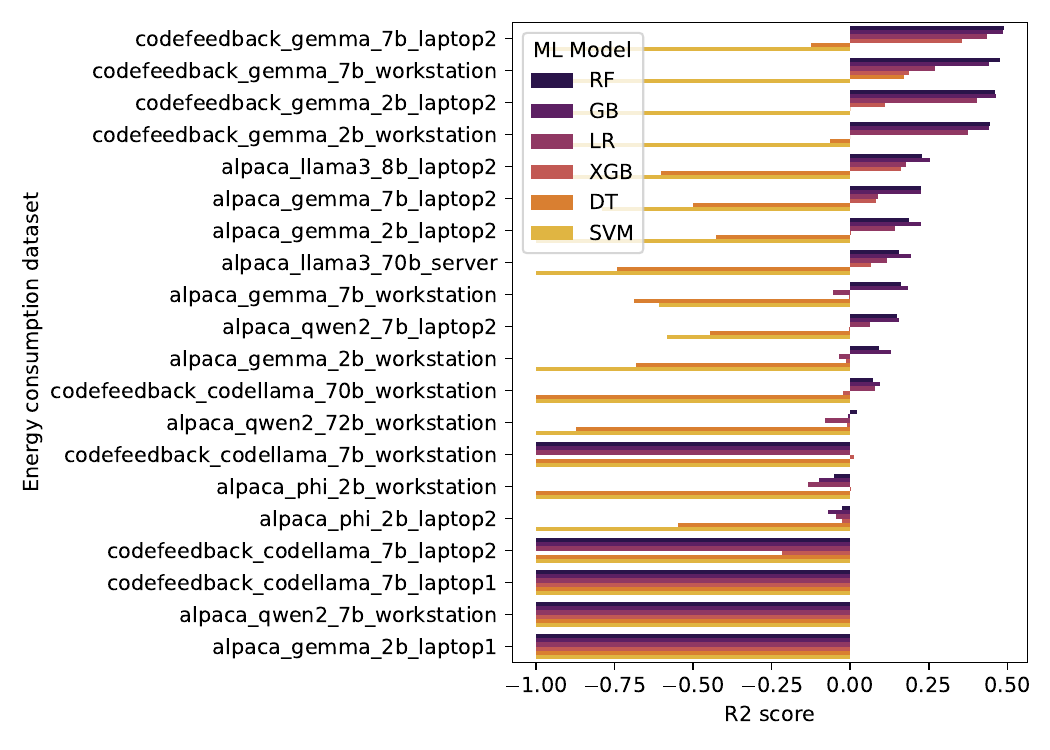}
    \vspace*{-1.9em}
    \caption{\response{R$^2$ scores across different ML algorithms, using prompt characteristics for forecasting energy consumption per response. Negative R$^2$ scores are clipped at -1.0.}}
    \label{fig:r2_ml_models_comparison}
    \vspace*{-0.8em}

\end{figure}

\begin{figure}[ht!]
    \centering
    \includegraphics[width=0.99\linewidth]{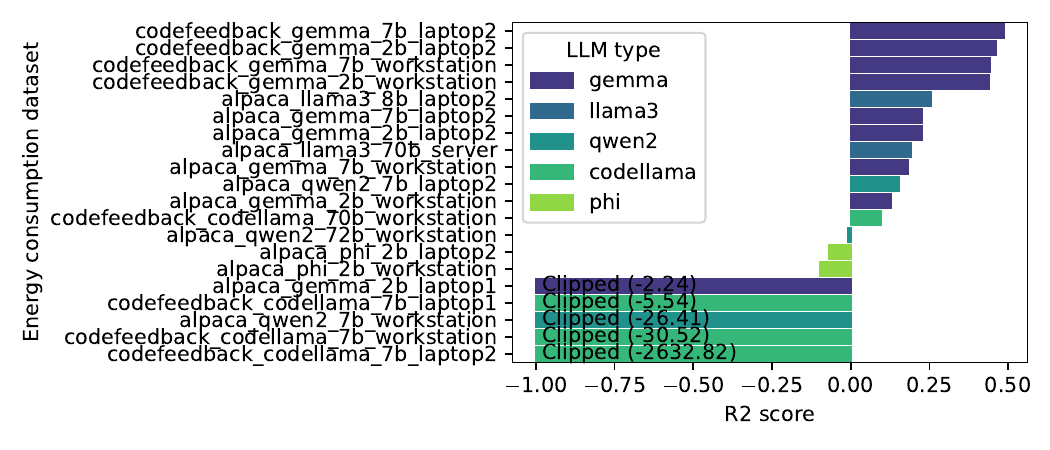}
    \vspace*{-1.60em}
    \caption{\response{R$^2$ scores of GB models using prompt characteristics for forecasting energy consumption per response.}}
    \label{fig:predictionscores}
    \vspace*{-1.50em}
\end{figure}

\response{Table~\ref{tab:correlations} shows that energy consumption is strongly correlated with response characteristics such as token length and duration, suggesting high potential for accurate prediction. 
Figure~\ref{fig:r2_ml_models_comparison_response} indicates that several ML models can predict energy consumption from response length with high accuracy: the best model achieves R$^2 > 0.78$ across all datasets, and R$^2 > 0.95$ for all but the bottom four. As summarized in Table~\ref{tab:r2scores}, LR performs best overall, with an average R$^2 = 0.96$ and top performance in 13 of 20 datasets. The lowest scores are observed on Laptop1 (R$^2 = 0.77$), indicating possible measurement inaccuracies in this CPU-only setup.}

Our results suggest that limiting response length (e.g., by specifying a maximum output size in the prompt) can substantially reduce inference energy consumption. In contrast, models trained solely on prompt-derived features show limited predictive performance (Figure~\ref{fig:predictionscores}).


%
\begin{table}[t]
    \centering
    \scriptsize
    \caption{\response{Average R$^2$ scores and number of wins for various ML algorithms using two feature sets: (i) only response length, and (ii) prompt characteristics.}}
    \label{tab:r2scores}
    \vspace*{-0.8em}
    \begin{tabular}{lrrrr}
        \toprule
        \multirow{2}{*}{ML alg.} & \multicolumn{2}{c}{Response length only} & \multicolumn{2}{c}{Prompt characteristics} \\
        \cmidrule(lr){2-3} \cmidrule(lr){4-5}
         & Avg. R$^2$ & No. of wins & Avg. R$^2$ & No. of wins \\
        \midrule
        DT  & 0.87      & 0  & -156.14   & 0 \\
        GB  & 0.89      & 5  & -134.72   & 10 \\
        LR  & 0.96      & 13 & -18.34    & 0 \\
        RF  & 0.89      & 2  & -48.01    & 6 \\
        SVM & -12392.08 & 0  & -12392.08 & 0 \\
        XGB & -1.17     & 0  & -0.47     & 4 \\
        \bottomrule
    \end{tabular}
    \vspace*{-1.4em}
\end{table}

\response{Figure~\ref{fig:r2_ml_models_comparison} reports R$^2$ scores for models trained solely on prompt-derived features, excluding any response information, to assess whether inference energy can be predicted a priori. Overall performance is low: only four datasets achieve R$^2 > 0.44$, and none exceed 0.50. As summarized in Table~\ref{tab:r2scores}, LR attains the highest average R$^2$, while GB performs best in roughly half of the datasets, indicating that predicting inference energy from prompt characteristics alone remains challenging. Notably, predictive accuracy varies substantially across datasets and appears primarily driven by model type: in Figure~\ref{fig:predictionscores}, GB models yield mostly positive R$^2$ for \texttt{Gemma}, low negative scores for \texttt{Phi}, and the worst performance for \texttt{CodeLlama}. In contrast, hardware, model size, and prompt dataset show no consistent effect. We hypothesize that model-specific training differences affect the predictability of response length from prompt features, but more controlled experiments are needed.} 



\begin{tcolorbox}[boxsep=2pt,left=2pt,right=2pt,top=2pt,bottom=2pt]



\response{\textbf{RQ3 Conclusion.} ML models, particularly Linear Regression, predict inference energy accurately when using response length as input, with most datasets achieving R$^2 > 0.95$. In contrast, models trained solely on prompt features perform poorly, though results vary widely across datasets. This suggests that some LLM families exhibit a stronger, more predictable relationship between prompt characteristics and response length than others.}

\end{tcolorbox}

\subsubsection{\response{RQ4 (mathematical model for energy consumption based on response length, model and hardware)}}


\response{The consistently high predictive performance of the LR models in RQ3 suggests that inference-time energy consumption can be approximated through a compact and interpretable mathematical formulation. To investigate this in RQ4, we aggregated data from all experiments and trained an LR model with response length, model attributes, and hardware configuration as predictors. We model the energy consumption $E$ (kWh) of a single inference instance as:}

\vspace*{-1.5em}

\begin{multline}
E = \beta_0 + \beta_1 \cdot n_{\text{tokens}} + \beta_2 \cdot s_{\text{model}} + \sum_{i} \alpha_i \cdot I_{\text{type}_i} + \\ \sum_{j} \gamma_j \cdot I_{\text{hw}_j} + \sum_{i,j} \delta_{i,j} \cdot (n_{\text{tokens}} \times I_{\text{cat}_{i,j}})
\end{multline}

\vspace*{-1.0em}


where:
\begin{itemize}
    \item $\beta_0$ is the base energy cost (intercept),
    \item $\beta_1$ is the baseline energy per response token,
    \item $n_{\text{tokens}}$ is the number of generated response tokens,
    \item $\beta_2$ is the coefficient for model size (parameters),
    \item $s_{\text{model}}$ is the model size,
    \item $\alpha_i$ are model type effects encoded as categorical indicators (codellama, gemma, llama3, phi, qwen2)
    \item $I_{\text{type}_i}$ are indicator variables for model types,
    \item $\gamma_j$ are hardware effects encoded as categorical indicators (laptop, workstation, server),
    \item $I_{\text{hw}_j}$ are indicator variables for hardware setups,
    \item $\delta_{i,j}$ are interaction coefficients that allow the energy-per-token rate to vary across model types and hardware configurations.
\end{itemize}

\response{Interaction terms were included because the empirical results indicate that token-level energy scaling differs across model architectures and hardware setups, making a single global slope insufficient to capture observed behavior. Numerical predictors include response length (primary driver) and model size, while categorical predictors include 11 models (\texttt{codellama}, \texttt{codellama-70b}, \texttt{codellama-7b}, \texttt{gemma-2b}, \texttt{gemma-7b}, \texttt{llama3}, \texttt{llama3-70b}, \texttt{phi}, \texttt{qwen2}, \texttt{qwen2-72b}, \texttt{qwen2-7b}) and 4 hardware configurations (Laptop 1, Laptop 2, Workstation, Server). We encoded categorical variables with one-hot encoding and removed the first category to avoid multicollinearity.}

\response{\textbf{Model Performance.} The fitted LR model achieves very strong performance across all combined datasets, reaching $R^2 = 0.9962$. This result indicates that inference energy can be predicted accurately using a simple, interpretable formulation grounded in response length, model identity, and deployment context.}

\response{\textbf{Estimated Coefficients and Interpretation.} The estimated baseline coefficient for response length was $\beta_1 = 5.28 \times 10^{-7}$ kWh/token, capturing the average marginal cost per generated token across all configurations. In contrast, model size exhibited only a negligible direct effect ($\beta_2 = 3.96 \times 10^{-18}$ kWh/billion parameters), suggesting that model size primarily impacts energy indirectly through \textbf{model-family and configuration-specific token-level scaling} captured by categorical and interaction terms.}

\response{Model family effects showed substantial variability in intercept shifts. For example:}

\vspace*{-0.3em}

\begin{itemize}

\item \texttt{codellama-70b} exhibited a strong negative intercept shift ($-5.67 \times 10^{-5}$ kWh - energy offset),

\item \texttt{qwen2-7b} showed the largest positive shift ($+3.98 \times 10^{-5}$),

\item \texttt{gemma} and \texttt{llama3} showed more moderate shifts in the range of $+2.4$ to $+3.2 \times 10^{-5}$ kWh.

\end{itemize}

\vspace*{-0.3em}

\response{Hardware-specific intercept terms differed, with \textbf{server} having the highest baseline energy, consistent with higher system-level power draw even for comparable workloads.}

\vspace*{-0.3em}

\begin{itemize}
    \item \textbf{Server:} $+3.24 \times 10^{-5}$ kWh (highest baseline cost, likely due to higher power draw)
    \item \textbf{Workstation:} $-1.42 \times 10^{-6}$ kWh (slight reduction in baseline cost)
    \item \textbf{Laptop 2:} $+2.31 \times 10^{-6}$ kWh (small positive shift)
\end{itemize}

\vspace*{-0.3em}

\response{More importantly, token-level scaling varied substantially across configurations. For example:}

\vspace*{-0.3em}

\begin{itemize}

\item \texttt{qwen2-72b} and \texttt{codellama-70b} showed the steepest energy-per-token scaling,

\item \texttt{llama3-70b} exhibited similarly high scaling,

\item Smaller models such as \texttt{gemma-2b} and \texttt{phi-2b} showed considerably lower scaling slopes.

\end{itemize}

\vspace*{-0.3em}

\response{These findings reinforce that inference energy is not determined solely by parameter count, but also by \textbf{architecture-dependent decoding behavior and hardware-dependent execution efficiency}.}

\response{\textbf{Ablation Study.} To quantify the relative impact of different feature groups, we performed an ablation study by fitting reduced LR models:}

\begin{itemize}
    \item Both interactions (full model): $R^2 = 0.9962$ 
    \item Hardware interactions only: $R^2 = 0.8361$ 
    \item Model type interactions only: $R^2 = 0.9954$ 
\end{itemize}

\response{These results indicate that \textbf{model type and response length explain most of the variance}, while hardware contributes a smaller but non-negligible portion. The findings suggest that, for practical energy-aware deployment, optimization should prioritize \textbf{selecting more efficient model families}, and \textbf{controlling response length}, while hardware choice plays a secondary role compared to these factors.}

\begin{tcolorbox}[boxsep=2pt,left=2pt,right=2pt,top=2pt,bottom=2pt]
\textbf{RQ4 Conclusion.} \response{The mathematical model demonstrates that LLM inference energy consumption is highly predictable (R$^2$ = 0.9962), with response length being the dominant factor and model type accounting for nearly all categorical variation. Our ablation study reveals that hardware configuration has minimal impact once model type is known, indicating that energy optimization should prioritize model selection and response length control over hardware choices.}


\end{tcolorbox}

%


\subsubsection{RQ5 (variability in energy consumption for the same prompt across multiple executions and different models)}

\begin{wrapfigure}{r}{.26\textwidth}
    \vspace*{-1.0em}
    \begin{minipage}{\linewidth}
    \centering
    \includegraphics[width=\linewidth]{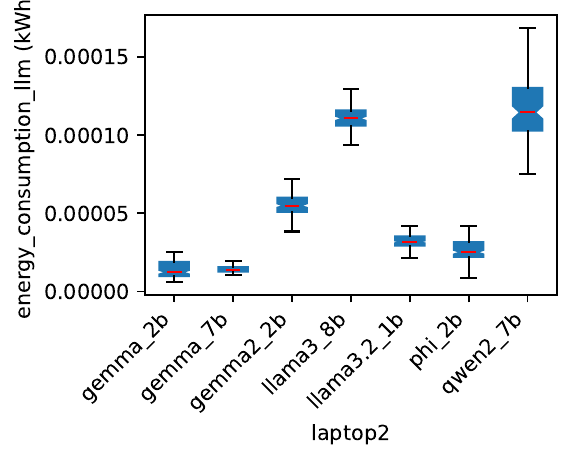}
     \vspace*{-2.0em}
    \subcaption{Variation in energy consumption per prompt for the same prompt run multiple times on seven models.}
    \label{fig:single_prompt_energy}\par\vfill
    \includegraphics[width=\linewidth]{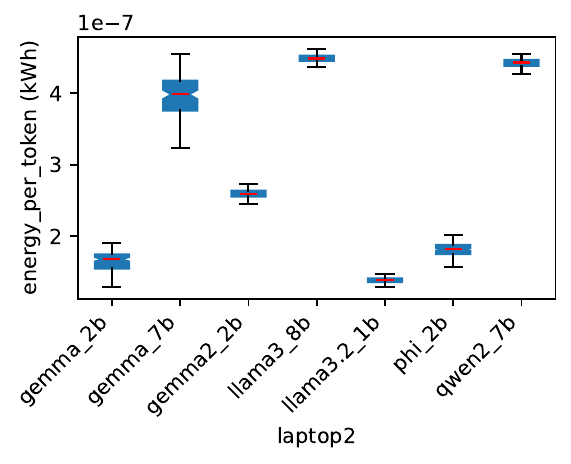}
    \vspace*{-2.0em}
    \subcaption{Variation in energy consumption per token for the same prompt run multiple times on seven models.}
    \label{fig:single_prompt_energy_per_token}
    \includegraphics[width=\linewidth]{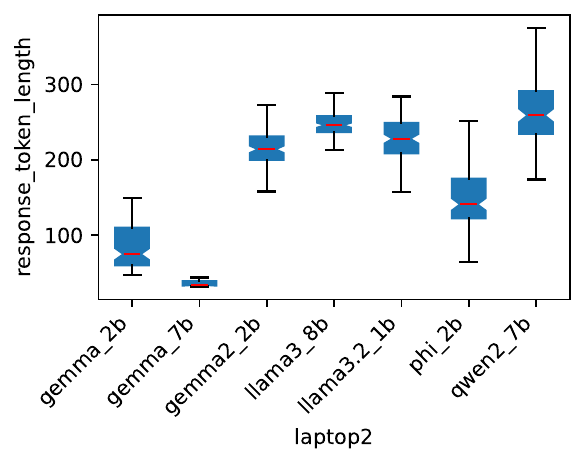}
    \vspace*{-2.0em}
    \subcaption{Variation in response token length for the same prompt run multiple times on seven models.}
    \label{fig:single_prompt_response_length}
\end{minipage}
\caption{Variability in energy consumption and response token length for 100 iterations of the prompt 'Give three tips for staying healthy' from \texttt{Alpaca}.} 
\label{fig:single_prompt}
\vspace*{-1.0em}
\end{wrapfigure}

To address RQ5, we analyzed the energy consumption and response token length for 100 iterations of a prompt (\textit{Give three tips for staying healthy}) from the \texttt{Alpaca} dataset on seven models using laptop 2. We evaluated variability by using the prompt across multiple trials. This ensures uniformity in our analysis, as the chosen prompt is unlikely to influence variations in energy consumption and response length when repeatedly processed. Figure~\ref{fig:single_prompt} presents differences in energy consumption and response variability among models. 


Figure~\ref{fig:single_prompt_energy} presents the energy consumption per prompt across seven models, revealing substantial variability. \texttt{Qwen2-7b}, the most expensive per prompt overall, exhibits the highest IQR of energy consumption. \texttt{Gemma-7b} exhibits consistently similar consumption per prompt as the smaller version of the model (\texttt{gemma-2b}). This aligns with its much lower response token length in Figure~\ref{fig:single_prompt_response_length}. Shorter responses generally reduce the total energy for processing. The variability is the smallest in all the models (see its IQR), suggesting similar energy use across repeated inferences.

Figure~\ref{fig:single_prompt_energy_per_token} displays the energy consumption per token, noting a wide IQR for \texttt{gemma-2b}, which is significantly greater than that of other models. While \texttt{gemma-7b} also shows a higher IQR, it is less pronounced than gemma-2b. The largest models (\texttt{gemma-7b}, \texttt{llama3-8b}, and \texttt{qwen2-7b}) exhibit the highest mean energy consumption per token, as they require more computation. This direct relationship between model size and energy expenditure is consistent with our findings, underscoring that the models with 7B and 8B parameters are more energy-intensive than their counterparts due to their complex inference processes.



\texttt{Llama3-8b} and \texttt{qwen2-7b} produce the longest responses, averaging 249 and 264 tokens, respectively (see Figure~\ref{fig:single_prompt_response_length}), contributing to their higher energy consumption per prompt. Despite their lengthy outputs, these models display moderate variability in response length, indicating consistent response generation for similar prompts.


\begin{tcolorbox}[boxsep=2pt,left=2pt,right=2pt,top=2pt,bottom=2pt]
\textbf{RQ5 Conclusion.} The results revealed significant differences in energy usage and response length variability among models. \texttt{Qwen2-7b} demonstrated larger fluctuations in energy consumption per prompt than others, suggesting factors such as model architecture and operational conditions may influence energy usage consistency. 

\end{tcolorbox}


\begin{table*}[ht]
\centering
\scriptsize
\caption{Comparison of energy measurement tools for LLM deployments.}
\begin{tabular}{lcccc}
\hline
\textbf{Criteria} & \textbf{MELODI} & \textbf{PyJoules} & \textbf{CodeCarbon} & \textbf{EnergyMeter} \\ \hline
\textbf{CPU Tracking} & RAPL (via Scaphandre) & RAPL & RAPL & RAPL (via pyRAPL) \\ 
\textbf{GPU Tracking} & NVIDIA-SMI (via NVML) & NVML & pynvml (NVML) & NVIDIA-SMI (via NVML) \\ 
\textbf{Default sampling freq.} & 10 Hz & Hardware-dependent & 1/15 Hz & Not specified \\ \hline
\end{tabular}
\label{tab:tool_comparison}
\vspace*{-1.0em}
\end{table*}

\subsubsection{RQ6 (comparative analysis of energy measurement tools in LLM deployments)}


To address RQ6, we evaluated per-token energy consumption using 467 to 480 prompts from the Alpaca dataset across four tools: \sysname, PyJoules~\cite{PyJoules}, CodeCarbon~\cite{codecarbon}, and EnergyMeter~\cite{argerich2024measuring}. We selected these tools for comparison due to their diverse methodologies in monitoring and reporting energy consumption in LLM deployments, allowing us to thoroughly assess and benchmark the performance of \sysname against established tools. We run \response{\texttt{phi2-2b}~\cite{phi2} on laptop 2 for the analysis. We opted for \texttt{phi2-2b}} primarily for its rapid inference capabilities. Given that our study prioritizes the assessment of measurement tools over model performance, utilizing a fast-processing model ensures uniform results across various tools and minimizes delays that could arise from complex model computations. The model choice is not expected to influence the observed differences between the monitoring tools, as the effects are normalized by averaging over the number of tokens.

\begin{figure}
    \centering
    \includegraphics[width=0.96\linewidth]{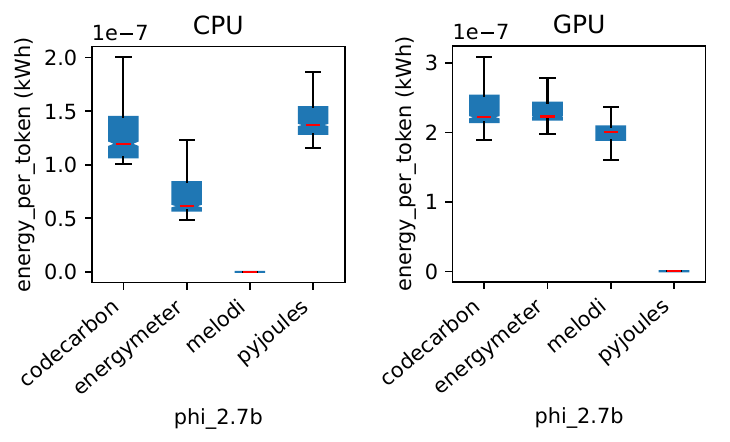}
    \caption{Comparison of the energy consumption per token, as measured by four different monitoring frameworks.}
    \label{fig:tool_comparison}
    \vspace*{-1.6em}
\end{figure}

Figure~\ref{fig:tool_comparison} displays considerable variation in per-token energy consumption across the tools. For CPU energy consumption, CodeCarbon recorded a mean of \(1.40 \times 10^{-7} \, \text{kWh}\), EnergyMeter logged \(8.13 \times 10^{-8} \, \text{kWh}\), and PyJoules reported \(1.46 \times 10^{-7} \, \text{kWh}\). Notably, the process-level measurements of \sysname showed the lowest average CPU energy at \(6.40 \times 10^{-11} \, \text{kWh}\). For GPU, CodeCarbon and EnergyMeter were closely aligned with means around \(2.57 \times 10^{-7} \, \text{kWh}\) and \(2.53 \times 10^{-7} \, \text{kWh}\), respectively. \sysname reported a mean GPU energy consumption of \(1.22 \times 10^{-7} \, \text{kWh}\), while PyJoules recorded a significantly lower mean GPU energy at \(2.81 \times 10^{-10} \, \text{kWh}\), possibly reflecting differences in how idle states or sampling frequencies impact measurements.


The variability in energy measurements among different tools suggests tool-specific biases or variations in interface handling for CPU and GPU measurements. For CPU, the per-process measurement approach of \sysname yielded significantly lower energy readings, potentially isolating energy consumption more effectively in the mixed-use environments of laptops. In contrast, PyJoules and CodeCarbon, which measure CPU energy at the system level, reported higher averages, likely including additional background activities that influence overall energy assessments.


The similar GPU energy consumption readings between CodeCarbon and EnergyMeter suggest similar NVML-based GPU data handling (Table~\ref{tab:tool_comparison}). Conversely, \sysname reported a slightly lower mean GPU energy consumption, indicating a more restrictive approach to GPU monitoring and potentially reflecting a more precise monitoring method. Notably, the lower IQR in \sysname suggests more accurate data collection, supported by its higher sampling frequency of 10 Hz compared to the lower frequency of CodeCarbon (Table~\ref{tab:tool_comparison}). PyJoules, having dramatically lower readings and outdated maintenance (its last update was in October 2021), raises concerns about its measurement accuracy. The variations in GPU energy measurements across tools highlight the potential impact of how each tool interacts with NVML-based monitoring on the reliability of GPU data.


\sysname distinguishes itself by its capability to measure CPU energy consumption at the per-process level, effectively isolating model-specific usage more accurately than whole-system methods. Its precision minimizes the interference from background activities prevalent in multitasking environments, enhancing its reliability. 

\begin{tcolorbox}[boxsep=2pt,left=2pt,right=2pt,top=2pt,bottom=2pt]
\textbf{RQ6 Conclusion.} The analysis of energy measurement tools for LLM deployments revealed significant differences in per-token energy consumption among \sysname, PyJoules, CodeCarbon, and EnergyMeter. \sysname's per-process CPU monitoring yielded more precise and generally lower energy readings compared to whole-system methods, pinpointing model-specific consumption. Moreover, the variability in GPU energy measurements, especially the abnormally low readings from PyJoules, suggests discrepancies that could stem from how the tools interact with NVIDIA’s GPU measurements, potentially affected by sampling frequencies and recording intervals.

\end{tcolorbox}

\subsection{Threats to validity}


\emph{\textbf{Internal Validity.}} 
The accuracy of the power usage monitoring tools significantly impacts the internal validity of our study. The nvidia-smi tool, according to NVIDIA documentation~\cite{nvidia_smi_docs}, has an accuracy of $\pm 5$ Watts, though Yang et al. suggest this error margin could be as high as $\pm 5\%$. Ensuring no concurrent GPU processes is crucial for accurate readings, and while most experiments isolated the GPU, residual base processes had minimal impact. Scaphandre’s reliance on Intel’s RAPL interface for power reasings may also affect data reliability. Variability in hardware power efficiency could further skew results, underscoring the need for more precise measurement tools and standardized testing across different hardware configurations in future research.

\emph{\textbf{External Validity.}} The generalizability of our findings is limited by the scope of our inference scenarios, particularly concerning the hardware setups and LLMs tested. Given the sample size, drawing broad conclusions is challenging. The ability to extend our results more widely hinges on whether the hardware and model configurations used accurately reflect the diversity in larger-scale LLM deployments.


\section{Discussion}
\label{sec:discussion}


\textbf{Scope of Our Study.} Our study concentrates on CPU and GPU energy consumption for LLM inference, chosen for its measurable impact. While significant, this focus omits other energy contributors in data center environments like networking, memory management, orchestration, and cooling, which can exceed computational energy use. This research addresses only direct CPU and GPU power use, emphasizing the need for broader energy evaluations in future work.

\textbf{Response Token Length as an Energy Indicator.} Our results reveal that response token length is a reliable indicator of energy consumption, suggesting that managing response lengths could effectively control energy use. 
However, the connection between prompt complexity and energy consumption is less definitive. While models based on prompt characteristics show potential, their performance varies, indicating these correlations may not be strong enough for reliable predictions in all contexts. 
We plan to explore the potential of using LLMs or other NLP models to analyze prompts for energy consumption prediction, balancing the benefits against the potential energy costs of such analyses.




\textbf{Exploring Response Quality and Energy Use.} Our current analysis does not address the quality of responses generated. Our dataset, however, provides a rich foundation to explore the relationship between response quality and energy consumption. Understanding whether higher-quality responses inherently require more energy could inform the development of more efficient models and prompt designs.


\textbf{Enhancing Generalizability and Model Comparisons.} To enhance our study's generalizability, we plan to test a wider array of hardware configurations and model types, allowing us to better understand energy consumption variations. Comparing the energy efficiency of task-specific models against general-purpose models in different applications would also be valuable. This analysis could reveal if specialized models yield significant energy savings for specific tasks, ensuring that comparisons maintain a consistent base model for accuracy.




\textbf{Investigating Energy Consumption Patterns.} Future research could explore patterns in energy consumption across models, focusing beyond mere size to include architectural differences. This could reveal how specific design choices impact energy demands and help optimize models for better energy efficiency without sacrificing performance.



\section{Related Work}
\label{sec:related_work}

\subsection{Energy and Carbon Footprint of AI and LLMs}


\response{The advancement of deep neural networks, particularly in natural language processing (NLP), has intensified concerns about their environmental impact. Strubell et al.~\cite{strubell2019energy} quantified the energy cost of training large NLP models, showing that a single training run can emit as much $\mathrm{CO_2}$ as multiple passenger vehicles over their lifetimes. Subsequent studies expanded this perspective: Luccioni et al.~\cite{luccioni2023estimating} analyzed inference-time energy consumption for large transformer models such as BLOOM, while  Luccioni et al.~\cite{luccioni2023power} demonstrated that general-purpose generative models incur orders-of-magnitude higher energy costs than task-specific alternatives, even when controlling for model size.}


\response{Recent work has shifted attention from training to the operational footprint of LLMs. De Vries~\cite{de2023growing} highlights the substantial contribution of AI-driven data center operations to global energy demand. Similarly, Jegham et al.~\cite{jegham2025hungry} examined energy, water, and carbon metrics across state-of-the-art LLM inference deployments, and Poddar et al.~\cite{poddar2025towards} identify strong correlations between inference energy, output length, model size, and hardware configuration. Meanwhile, Google Cloud’s analysis of their Gemini model estimate median prompt energy usage at 0.24 W·h and $\mathrm{CO_2}$ emission at 0.03 g, underscoring how inference, though individually modest, scales massively~\cite{google_blog}.}


\response{Collectively, this body of work underscores a critical insight: while training remains energy-intensive, \textbf{inference increasingly dominates the lifecycle energy and carbon footprint of LLMs} due to continuous, large-scale deployment. Despite this shift, most prior studies still emphasize training or rely on coarse-grained measurements, motivating the need for precise, deployment-aware methodologies to analyze inference-time energy consumption and support sustainable AI engineering.}

\subsection{Energy Monitoring Frameworks for LLMs}


\response{Prior work has proposed several frameworks for monitoring and analyzing the energy use and carbon footprint of LLMs, revealing substantial variation across model configurations~\cite{liang2022holistic, samsi2023words, wilkins2024offline, wilkins2024hybrid, argerich2024measuring}. Wilkins et al.~\cite{wilkins2024offline} introduce an offline, workload-based framework for LLM inference on heterogeneous CPU–GPU systems, modeling energy consumption and runtime as functions of input and output tokens. However, their reliance on AMD$\mu$Prof entails notable limitations, including measurement overhead from frequent polling, limited real-time granularity, and hardware specificity to AMD CPUs, which restricts generalizability and increases setup complexity. In contrast, Samsi et al.~\cite{samsi2023words} focus exclusively on GPU energy consumption for LLaMA models, omitting CPU contributions.}

\response{Our work complements these efforts above by providing a more comprehensive framework that measures energy consumption across multiple LLMs, using Scaphandre and nvidia-smi for real-time tracking of both CPU and GPU energy usage. This broader approach enables a more holistic understanding of energy dynamics during LLM inference.} 


\response{Argerich et al.~\cite{argerich2024measuring} present a related framework based on pyRAPL~\cite{pyRAPL}, nvidia-smi~\cite{nvidia_smi_docs}, and eBPF~\cite{eBPF} to estimate CPU, memory, GPU, and storage energy on Linux, supported by extensive analyses across many LLMs and model sizes. In contrast, \sysname supports a broader range of hardware configurations beyond a single bare-metal environment and extends energy monitoring with predictive analysis, providing a more flexible and comprehensive framework for studying LLM inference energy consumption. 
Henderson et al.~\cite{henderson2020towards} utilize RAPL in a similar manner to Scaphandre to monitor energy use but derive per-process energy consumption based on the CPU's relative utilization. Scaphandre, however, provides a more comprehensive explanation of how it calculates per-process energy metrics.}

\subsection{Inference-Time Energy Profiling and Optimization}


\response{The growing deployment of LLMs in production highlights the importance of inference-time energy profiling and optimization. Samsi et al.~\cite{samsi2023words} analyze inference energy costs for different LLaMA model sizes on NVIDIA V100 and A100 GPUs, demonstrating how model scale, hardware, and task characteristics influence energy consumption. Similarly, Fernandez et al.~\cite{fernandez2025energy} show that inference optimizations—such as quantization, batching strategies, and hardware-aware mapping—can reduce energy usage by up to 73\% compared to unoptimized deployments. Wilhelm et al.~\cite{wilhelm2025beyond} introduce the energy-per-token metric and demonstrate strong correlations between response length, model size, and energy consumption, while also exploring dynamic reasoning strategies to reduce energy without significant accuracy loss. Stojkovic et al.~\cite{stojkovic2025dynamollm} introduce a dynamic inference-cluster management framework that exploits workload heterogeneity to optimize inference energy, achieving average energy savings of 52\% under service-level objectives.}

\response{Despite these advances, most existing approaches lack fine-grained attribution of energy usage to specific processes (e.g., model decoding) or integrations across diverse hardware and prompt workloads. This gap motivates frameworks capable of capturing per-token, per-process, and hardware-aware energy profiles for real-world LLM deployment scenarios.}

\subsection{Hardware-Level Power Monitoring Techniques}


\response{Hardware-level power monitoring is fundamental for analyzing and optimizing energy consumption in computing systems. Intel’s RAPL (Running Average Power Limit) interface is widely used, providing on-chip energy counters for CPU and DRAM domains. Prior studies show that RAPL measurements strongly correlate with external power meters, although limitations remain due to driver support, update timing, and domain-specific inaccuracies~\cite{khan2018rapl}. In particular, RAPL’s memory-domain estimates can overestimate power consumption and exhibit temporal inconsistencies on heterogeneous memory systems~\cite{alt2024experimental}, highlighting the need for careful interpretation of such measurements.}

\response{On the GPU side, power monitoring commonly uses NVML (NVIDIA Management Library) and the nvidia‑smi interface. You et al.~\cite{you2023zeus}} 
\response{show that GPU power limits strongly affect the energy–performance trade-off of DNN workloads.
Another study by Yang et al.~\cite{yang2023part} reveals that nvidia-smi often samples only a portion of runtime, leading to substantial under- or over-estimation of energy use—sometimes by as much as 35–65\%. Furthermore, Arafa et al.~\cite{arafa2020verified} demonstrate how hardware counters and instrumentation can measure energy consumption at the instruction level on NVIDIA GPUs, reinforcing that architecture-specific factors (e.g., micro-architecture generation, instruction type) influence energy behavior significantly.}

\response{In summary, hardware-level measurement techniques provide essential visibility into component-level power usage (CPU, DRAM, GPU), but each comes with its own limitations—granularity, accuracy, sampling frequency, and applicability vary. For energy-aware systems especially in the context of LLM deployment, selecting and validating these techniques is critical before interpreting results.}

\subsection{Sustainable AI Engineering and Future Directions}



\response{Recent work on sustainable AI increasingly emphasizes energy- and resource-aware design, deployment, and operation of large models. A comprehensive survey on Green AI~\cite{bolon2024review} highlights the need to account for lifecycle impacts, including carbon and water usage, when evaluating AI systems. Another study on LLM sustainability~\cite{singh2025survey} extend this perspective by considering economic and resource costs alongside computation, advocating greater transparency and standardized benchmarking in AI energy and carbon accounting. Fernandez et al.~\cite{fernandez2025energy} show that inference-level optimizations can reduce energy consumption by up to 73\% with negligible impact on accuracy. Similarly, Sánchez-Mompó et al.~\cite{sanchez2025green} empirically demonstrate that larger models do not necessarily incur higher energy per request, particularly under low serving utilization.}


\response{Collectively, these studies underscore that sustainable AI engineering demands a holistic perspective encompassing model design, hardware selection, inference architecture, measurement transparency, and deployment practices. They point to key directions such as standardized efficiency metrics, energy-aware architectures, cross-layer co-design, and system-level optimizations. \sysname contributes to this agenda by providing fine-grained inference-time energy measurements supporting energy-conscious AI systems.}

\subsection{Research Gaps and Motivation for Our Framework}

\response{Although significant advances have been made in quantifying the environmental footprint of artificial intelligence systems, several research gaps remain—especially regarding \textbf{fine-grained, inference-time energy measurement of LLMs}. Current literature and tools leave important methodological, technical, and analytical challenges unresolved, motivating the need for a more comprehensive framework.}

\response{\textbf{(1) Lack of inference-level granularity.} Most prior research emphasizes training-phase emissions and provides only coarse-grained reporting for inference. Existing tools such as \textbf{CodeCarbon}, \textbf{PyJoules}, and \textbf{EnergyMeter} measure energy at the system level, aggregating consumption across all active processes. This approach obscures the specific energy cost of the LLM process, especially in multi-tasking environments, thereby limiting the interpretability of results and their use for optimization.} 

\response{\textbf{(2) Hardware-dependent and non-standardized monitoring.} Many tools rely on vendor-specific interfaces (e.g., Intel’s RAPL for CPUs and NVIDIA’s NVML for GPUs) whose accuracy, update frequency, and calibration differ across architectures. These inconsistencies introduce measurement bias and hinder the reproducibility and comparability of studies. Moreover, most frameworks do not jointly monitor CPU and GPU energy, overlooking cross-component dependencies shaping overall power dynamics.}

\response{\textbf{(3) Limited linkage between operational factors and energy behavior.} Empirical analyses rarely explore how prompt features (e.g., length, complexity), model configurations (e.g., size, architecture), or deployment hardware collectively shape energy consumption. This lack of fine-grained correlation hinders the development of targeted optimization strategies for sustainable inference.}


\response{\textbf{(4) Absence of unified, open, and process-level measurement solutions.} While recent works call for transparency and standardized reporting of energy metrics, there is no openly available framework that provides process-level monitoring, configurable sampling, and reproducible data collection across heterogeneous hardware setups.}

\response{\textbf{Rationale.} To address these challenges, we propose an extensible framework for high-resolution monitoring and analysis of LLM inference energy consumption. \sysname is designed to (i) capture CPU and GPU energy at the process level with configurable sampling intervals; (ii) log inference metadata, including token counts and timestamps; and (iii) enable empirical studies linking energy use to model, prompt, and hardware characteristics. By bridging measurement accuracy, analytical depth, and reproducibility, our framework advances the methodological foundations of sustainable AI engineering, enabling more transparent and data-driven optimization of LLM inference efficiency.}

\section{Conclusion}
\label{sec:conclusion}

In conclusion, \sysname represents a noteworthy leap forward in advancing energy-aware practices within the domain of LLMs. By formulating a precise methodology and unveiling an open-source instrument tailored for real-time monitoring of energy consumption throughout the LLM inference process, \sysname effectively bridges a vital chasm in sustainable computing. Our data collection, encompassing various LLMs, hardware setups, and prompt datasets, provides a robust platform for rigorous comparison and nuanced analysis. This facilitates a deeper understanding of the energy dynamics associated with LLM deployment scenarios, offering critical insights that contribute to the sustainability of LLMs.



\section*{Acknowledgment}
The work has been conducted as part of the ENFIELD project (101120657) funded by the European Commission within the HEU Programme. 

\bibliographystyle{IEEEtran}
\bibliography{references}

\begin{IEEEbiography}[{\includegraphics[width=1in,height=1.25in,clip,keepaspectratio]{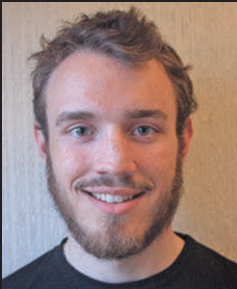}}]{Erik Johannes Husom} is a Research Scientist in the
Trustworthy Green IoT Software research group at SINTEF Digital, 0373 Oslo, Norway. His research interests include applied ML, AI engineering, and responsible AI. Husom received
his M.Sc. in computational physics from the University of Oslo. 
\end{IEEEbiography}

\vspace*{-3.0em}

\begin{IEEEbiography}[{\includegraphics[width=1in,height=1.25in,clip,keepaspectratio]{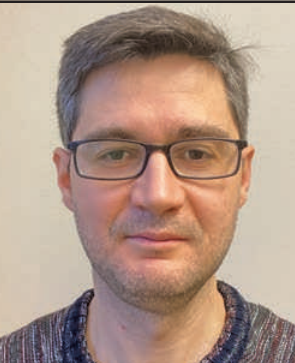}}]{Arda Goknil} received his Ph.D. degree in computer science from the University of Twente, in the Netherlands, in 2011. He is a senior research scientist at SINTEF, Norway. He was a research associate at the University of Luxembourg. His research concerns AI Engineering, Sustainable AI, Software Testing, Software Security, and Intermittent Computing. He is active on EU-funded and national research projects with several academic and industry partners.
\end{IEEEbiography}


\vspace*{-3.0em}

\begin{IEEEbiography}[{\includegraphics[width=1in,height=1.25in,clip,keepaspectratio]{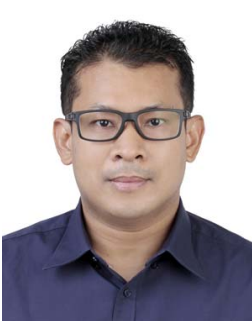}}]{Lwin Khin Shar} (Member, IEEE) received a Ph.D. degree in software engineering from Nanyang Technological University (NTU), Singapore, in 2014. He is an Associate Professor at the School of Computing and Information Systems, Singapore Management University, Singapore. He was a Postdoctoral Research Associate with SnT of the University of Luxembourg, Esch-sur-Alzette, Luxembourg, and then a Research Scientist with NTU. His research interests include software engineering, security \& privacy, and machine learning. 

\end{IEEEbiography}

\vspace*{-3.0em}

\begin{IEEEbiography}[{\includegraphics[width=1in,height=1.25in,clip,keepaspectratio]{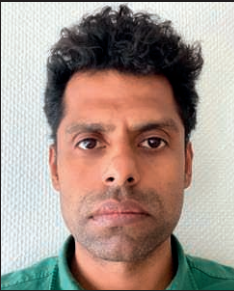}}]{Sagar Sen} is a senior research scientist at SINTEF Digital, 0373 Oslo, Norway. His research interests are in engineering and testing of lifelong AI systems for application domains
such as manufacturing and health. Sen received his Ph.D. in computer science from the University of Rennes 1/INRIA. He is a Member of IEEE. 
\end{IEEEbiography}

\end{document}